\documentclass[%
 reprint,
superscriptaddress,
 amsmath,amssymb,
 aps, singlecolumn
]{revtex4-1}

\usepackage[english,ngerman]{babel}
\usepackage{changes}
\usepackage{graphicx}% Include figure files
\usepackage{dcolumn}% Align table columns on decimal point
\usepackage{bm}% bold math
\PassOptionsToPackage{english}{babel}
\begin{document}

% TJK: please dont remove title suggestions, only comment them out.
%\title{Microresonator Soliton Frequency Combs for Ultrafast Optical Distance Metrology}
\title{Ultrafast optical ranging using microresonator soliton frequency combs}
%\title{Ultrafast distance measurements using photonic chip based soliton frequency combs}
 
\author{Philipp~Trocha}
\thanks{D.G., P.T., and M.K. contributed equally to this work}
\affiliation{Institute of Photonics and Quantum Electronics (IPQ), Karlsruhe Institute of Technology (KIT), 76131 Karlsruhe, Germany}
 
\author{Denis~Ganin}
\thanks{D.G., P.T., and M.K. contributed equally to this work}
\affiliation{Institute of Photonics and Quantum Electronics (IPQ), Karlsruhe Institute of Technology (KIT), 76131 Karlsruhe, Germany}

\author{Maxim~Karpov}
\thanks{D.G., P.T., and M.K. contributed equally to this work}
\affiliation{{\'E}cole Polytechnique F{\'e}d{\'e}rale de Lausanne (EPFL), CH-1015 Lausanne,
Switzerland}

\author{Martin~H.\,P.\,Pfeiffer}
\affiliation{{\'E}cole Polytechnique F{\'e}d{\'e}rale de Lausanne (EPFL), CH-1015 Lausanne,
Switzerland}

\author{Arne~Kordts}
\affiliation{{\'E}cole Polytechnique F{\'e}d{\'e}rale de Lausanne (EPFL), CH-1015 Lausanne,
Switzerland}

\author{Jonas~Krockenberger}
\affiliation{Institute of Photonics and Quantum Electronics (IPQ), Karlsruhe Institute of Technology (KIT), 76131 Karlsruhe, Germany}

\author{Stefan~Wolf}
\affiliation{Institute of Photonics and Quantum Electronics (IPQ), Karlsruhe Institute of Technology (KIT), 76131 Karlsruhe, Germany}

\author{Pablo~Marin-Palomo}
\affiliation{Institute of Photonics and Quantum Electronics (IPQ), Karlsruhe Institute of Technology (KIT), 76131 Karlsruhe, Germany}

\author{Claudius~Weimann}
\affiliation{Institute of Photonics and Quantum Electronics (IPQ), Karlsruhe Institute of Technology (KIT), 76131 Karlsruhe, Germany}
\affiliation{Now with: Corporate Research and Technology, Carl Zeiss AG, Oberkochen, Germany}

\author{Sebastian~Randel}
\affiliation{Institute of Photonics and Quantum Electronics (IPQ), Karlsruhe Institute of Technology (KIT), 76131 Karlsruhe, Germany}
\affiliation{Institute of Microstructure Technology (IMT), Karlsruhe Institute of Technology (KIT), 76344 Eggenstein-Leopoldshafen}

\author{Wolfgang~Freude}
\affiliation{Institute of Photonics and Quantum Electronics (IPQ), Karlsruhe Institute of Technology (KIT), 76131 Karlsruhe, Germany}
\affiliation{Institute of Microstructure Technology (IMT), Karlsruhe Institute of Technology (KIT), 76344 Eggenstein-Leopoldshafen}

\author{Tobias~J.~Kippenberg}
\email[]{tobias.kippenberg@epfl.ch}
\affiliation{{\'E}cole Polytechnique F{\'e}d{\'e}rale de Lausanne (EPFL), CH-1015 Lausanne,
Switzerland}

\author{Christian~Koos}
\email[]{christian.koos@kit.edu}
\affiliation{Institute of Photonics and Quantum Electronics (IPQ), Karlsruhe Institute of Technology (KIT), 76131 Karlsruhe, Germany}
\affiliation{Institute of Microstructure Technology (IMT), Karlsruhe Institute of Technology (KIT), 76344 Eggenstein-Leopoldshafen}

\date{\today}% It is always \today, today,

\begin{abstract}
    Light detection and ranging (LIDAR) is critical to many fields in science and industry. Over the last decade, optical frequency combs were shown to offer unique advantages in optical ranging, in particular when it comes to fast distance acquisition with high accuracy. However, current comb-based concepts are not suited for emerging high-volume applications such as drone navigation or autonomous driving. These applications critically rely on LIDAR systems that are not only accurate and fast, but also compact, robust, and amenable to cost-efficient mass-production. Here we show that integrated dissipative Kerr-soliton (DKS) comb sources provide a route to chip-scale LIDAR systems that combine sub-wavelength accuracy and unprecedented acquisition speed with the opportunity to exploit advanced photonic integration concepts for wafer-scale mass production. In our experiments, we use a pair of free-running DKS combs, each providing more than 100 carriers for massively parallel synthetic-wavelength interferometry. We demonstrate dual-comb distance measurements with record-low Allan deviations down to 12\,nm at averaging times of 14\,$\mu$s as well as ultrafast ranging at unprecedented measurement rates of up to 100\,MHz. We prove the viability of our technique by sampling the naturally scattering surface of air-gun projectiles flying at 150\,m/s (Mach 0.47). Combining integrated dual-comb LIDAR engines with chip-scale nanophotonic phased arrays, the approach could allow widespread use of compact ultrafast ranging systems in emerging mass applications.
\end{abstract}

%The ability to measure precise and fast distances using multi-wavelength based distance ranging using lasers (LIDAR) 

\maketitle

%\tableofcontents

%%%%%%%%%%%%%%%%%%%%%%%%%%%%%%%%%%%%%%%%%%%%%%%%
%%%%%%%%%%%%%%%%%%%%%%%%%%%%%%%%%%%%%%%%%%%%%%%%
\subsection*{Introduction}
Laser-based light detection and ranging (LIDAR) is widely used in science and industry, offering unique advantages such as high precision, long range, and fast acquisition \cite{amann2001laserRangingReview,Berkovic2012}. Over the last decades, LIDAR systems have found their way into a wide variety of applications, comprising, e.g., industrial process monitoring \cite{park1994opticalMeasSpindel}, autonomous driving \cite{levinson2011towardsAutonomousDriving}, satellite formation flying \cite{sassen2008clouds2Lidars}, or drone navigation \cite{li2014lidarDrone}. When it comes to fast and accurate ranging over extended distances, optical frequency combs \cite{Udem2002} have been demonstrated to exhibit unique advantages, exploiting time-of-flight (TOF) schemes \cite{minoshima2000combLIDAR}, interferometric approaches \cite{schuhler2006comb2wavelength}, or combinations thereof \cite{Coddington2009}. In early experiments by Minoshima \emph{et al.} \cite{minoshima2000combLIDAR}, mode-locked fiber lasers were used for TOF ranging, thereby primarily exploiting the stability of the repetition rate. Regarding interferometric schemes, optical frequency combs were exploited to stabilize the frequency interval between continuous-wave (CW) lasers used in synthetic-wavelength interferometry \cite{schuhler2006comb2wavelength,Jang.2016}. Dual-comb schemes, which rely on multi-heterodyne detection by coherent superposition of a pair of slightly detuned frequency combs, allow to combine TOF measurements with optical interferometry, thereby simultaneously exploiting the radio frequency coherence of the pulse train and the optical coherence of the individual comb tones \cite{Coddington2009}. More recently, comb-based schemes have been demonstrated as a viable path to high-speed sampling with acquisition times down to 500\,ns \cite{Ataie2013}.

However, besides accuracy and acquisition speed, footprint has become an important metric for LIDAR systems, driven by emerging high-volume applications such as autonomous drone or vehicle navigation, which crucially rely on compact and lightweight implementations, and by recent advances in photonic integration showing that large-scale nanophotonic phased arrays  \cite{schuhler2006comb2wavelength,doylend20112Dbeamsteering,sun2013nanophotonicPhasedArrray,Hulme2015} open a promising path towards ultra-compact systems for rapid high-resolution beam steering. To harness the full potential of these approaches, the optical phased arrays need to be complemented by chip-scale LIDAR engines that combine high precision with ultrafast acquisition and that are amenable to efficient mass production. Existing dual-comb LIDAR concepts cannot fulfill these requirements since they either rely on cavity-stabilized mode-locked fiber lasers\,\cite{Coddington2009} or on spectral broadening of initially narrowband seed combs\,\cite{Ataie2013}, which typically requires delicate fiber-based dispersion management schemes, usually in combination with intermediate amplifiers. These approaches are not suited for chip-scale photonic integration.

%. As a consequence, chip-scale mass-producible LIDAR engines with ultra-high sampling rates appear as a eminent technology gap. 

%However, all these demonstrations have in common that they either rely on cavity-stabilized mode-locked fiber lasers \cite{Coddington2009} or on spectral broadening of initially narrowband seed combs, which typically requires delicate dispersion management schemes, usually in combination with intermediate amplifiers \cite{Ataie2013}. These approaches are not easily amenable to photonic integration, and, in the case of fiber lasers, have limited acquisition speed. As a consequence, chip-scale mass-producible LIDAR engines with ultra-high sampling rates appear as a major technology gap, in particular in view of emerging LIDAR applications such as autonomous drones or vehicles, which crucially rely lightweight systems for fast and robust ranging to map surrounding objects for safe, weather-independent navigation. This need is emphasized even more by the advent of large-scale nanophotonic phased arrays  \cite{doylend20112Dbeamsteering,schuhler2006comb2wavelength} that promise a compact and reliable solution for rapid high-resolution beam steering.

%in particular in view of emerging LIDAR applications such as autonomous drones or vehicles, which crucially rely lightweight systems for fast and robust ranging to map surrounding objects for safe, weather-independent navigation. 

In this paper we show that integrated dissipative Kerr soliton \cite{akhmediev2005dissipative,Herr2013} (DKS) comb sources provide a route to chip-scale LIDAR systems that combine sub-wavelength accuracy and unprecedented acquisition speed with the ability to exploit advanced photonic integration concepts for cost-efficient mass manufacturing. DKS comb generators rely on high-\emph{Q} microresonators that can be efficiently fabricated in large quantities and stand out due to both large optical bandwidth and large free spectral range (FSR). DKS combs have previously been used in dual-comb spectroscopy \cite{dualcomb2016vahala,yu2016DKSdualcombMIR}, coherent communications \cite{Marin-Palomo2016}, and frequency metrology \cite{brasch2017DKSselfreferencing, Jost2014link}. In our demonstration, we use a pair of free-running DKS combs, each providing more than 100 carriers for massively parallel synthetic-wavelength interferometry. The large optical bandwidth of more than 11\,THz leads to highly precise distance measurements with record-low Allan deviations reaching 12\,nm at an averaging time of 14\,$\mu$s, while the large FSR enables high-speed measurements at rates of up to 100\,MHz. This is the highest measurement rate achieved in any optical distance measurement experiment so far, exceeding the fastest previous demonstration \cite{Ataie2013} by more than one order of magnitude. In contrast to previous dual-comb experiments \cite{Coddington2009, Ataie2013}, our scheme is based on a pair of free-running comb generators and does not require phase-locking of the combs to each other. We prove the viability of our technique by sampling the naturally scattering surface of air-gun projectiles on the fly, achieving lateral spatial resolutions of better than 2\,$\mu$m for object speeds of more than 150\,m/s, i.e., Mach 0.47. We expect that DKS-based dual-comb LIDAR could have transformative impact on all major application fields that require compact LIDAR systems and high precision ranging, in particular when combined with large-scale nanophotonic phased arrays \cite{sun2013nanophotonicPhasedArrray,doylend20112Dbeamsteering,schuhler2006comb2wavelength}. Acquisition rates of hundreds of megahertz could enable ultrafast 3D imaging with megapixel resolution and hundreds of frames per second.

\begin{figure*}
\includegraphics[width = \textwidth]{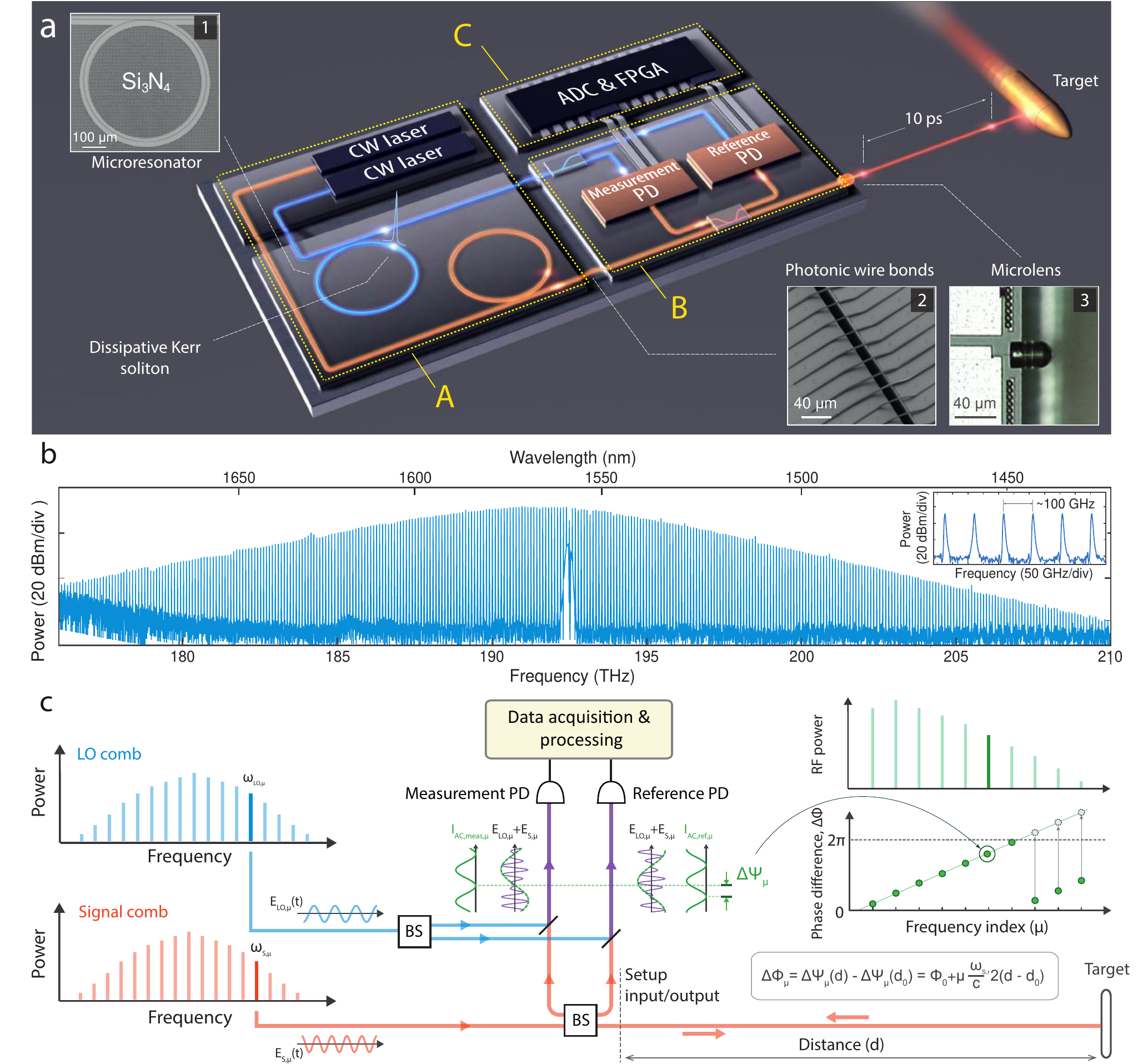}
\protect\caption{\textbf{\boldmath Vision of a chip-scale LIDAR system for ultrafast optical ranging.}
\textbf{(a)} Artist's view of a dual-comb chip-scale LIDAR engine. The system consists of a dual frequency comb source (A), a photonic integrated circuit (PIC) for transmission and detection of the LIDAR signal (B), as well as data acquisition and signal processing electronics (C). For comb generation, light of two continuous-wave (CW) pump lasers is coupled to the silicon nitride ($\rm{Si_{3}N_{4}}$) microresonators of the dual-comb source \cite{Pfeiffer2015Damascene}, Inset\,(1), where dissipative Kerr solitons (DKS) with broadband smooth spectra are generated via four-wave mixing processes \cite{Brasch2014Cherenkov}. All photonic integrated circuits are connected by photonic wire bonds, Inset\,(2) \cite{Lindenmann2012}. One of the two DKS combs is used for measuring the distance to the target ("signal comb", orange), whereas the other comb acts as a local oscillator ("LO comb", blue) for multi-heterodyne detection on balanced photodetectors (PD). On the LIDAR PIC, the signal comb is split in two parts. One part is collimated by a chip-attached micro-lens, Inset\,(3), sent to the measurement target, and the scattered light is coupled back into the on-chip waveguide and superimposed with a first portion of the LO comb in the measurement photodetector (Measurement PD). The other part of the signal comb is directly guided to the reference photodetector (Reference PD) along with the other portion of the LO comb. Distance information is extracted from the electrical beat notes in the photocurrents by a combination of analog-to-digital converters (ADC) and a field-programmable gate array (FPGA). \textbf{(b)} Spectrum of a DKS comb, featuring a smooth envelope and a line spacing of approximately 100 GHz. \textbf{(c)} Measurement principle: The signal comb (orange) and the LO comb (blue) consist of discrete tones at frequencies $\omega_{\rm{S},\mu}$ and  $\omega_{\rm{LO},\mu}$. The superposition of these comb lines on the measurement and reference PD leads to a multitude of beat notes in the RF spectrum of the photocurrent (green) at frequencies  $\Delta\omega_{µ} = \left|\omega_{\rm{LO,}\mu} - \omega_{\rm{S,}\mu}\right|$, which can be separated by a Fourier transformation. The phases of these beat notes reveal the phase shifts $\Psi_{\rm{meas},\mu}$ and $\Psi_{\rm{ref,\mu}}$ that the optical waves have accumulated. The distance is finally obtained by estimating the slope of the phase differences $\Delta\Phi_{\mu}$ according to Eq.~(\ref{Eq.2}) as a function of frequency index $\mu$, see lower right-hand side of Fig.\,\ref{fig_1}c. A detailed mathematical explanation can be found in the Supplementary Information.}
\label{fig_1}
\end{figure*}

%%%%%%%%%%%%%%%%%%%%%%%%%%%%%%%%%%%%%%%%%%%%%%%%
%%%%%%%%%%%%%%%%%%%%%%%%%%%%%%%%%%%%%%%%%%%%%%%%
\section*{Vision and concept}
%%%%%%%%%%%%%%%%%%%%%%%%%%%%%%%%%%%%%%%%%%%%%%%%
%%%%%%%%%%%%%%%%%%%%%%%%%%%%%%%%%%%%%%%%%%%%%%%%

The vision of an ultrafast chip-scale DKS-based LIDAR system is illustrated in Fig.\,\ref{fig_1}a. The system is realized as an optical multi-chip assembly combining different photonic integrated circuits (PIC) such as continuous-wave (CW) pump lasers, microresonator Kerr comb generators, Fig.\,\ref{fig_1}a - A, and a dedicated distance measurement chip, Fig.\,\ref{fig_1}a - B \cite{Weimann2014}, which contains data acquisition and signal processing electronics, Fig.\,\ref{fig_1}a - C. DKS combs  \cite{Brasch2014Cherenkov, Pfeiffer2015Damascene} are generated by a pair of high-\emph{Q} silicon nitride ($\rm{Si}_{3}\rm{N}_{4}$) microresonators, see Inset\,(1) of Fig.\,\ref{fig_1}a, which are pumped by III-V semiconductor CW lasers fabricated on a separate chip and coupled to the $\rm{Si}_{3}\rm{N}_{4}$ chip by photonic wire bonds (PWB), see Inset\,(2) of Fig.\,\ref{fig_1}a \cite{Lindenmann2012,Billah2017}. DKS combs combine a broadband optical spectrum having a smooth envelope with large line spacings of, e.g., $100\,\rm{GHz}$, see Fig.\,\ref{fig_1}b. One of the two DKS combs is used to probe the distance to the target ("signal comb", depicted in orange), whereas the other comb acts as a local oscillator ("LO comb", depicted in blue) for massively parallel multi-heterodyne detection on a pair of balanced photodetectors (PD). The signal comb is split in two parts. One part is collimated by a chip-attached microlens, see Inset\,(3) of Fig.\,\ref{fig_1}a \cite{Dietrich2016}, sent to the measurement target, and the backscattered light is coupled back into the on-chip waveguide and superimposed with a first portion of the LO comb in the measurement detector (Measurement PD) for multi-heterodyne detection. The other part of the signal comb is directly guided to the reference detector (Reference PD).  
%to generate a phase reference for extracting the distance information, see Supplementary Information for details. 
Signal processing and distance extraction is accomplished by a combination of high-speed analog-to-digital converters (ADC) and a field-programmable gate array (FPGA). 

The measurement principle combines synthetic-wavelength interferometry with massively parallel comb-based multi-heterodyne detection and is illustrated in Fig.\,\ref{fig_1}c. In the following, we only give a brief outline of the technique -- please refer to the Supplementary Information for a rigorous mathematical description. For our analysis, we use a space and time dependence of optical waves of the form $\textrm{exp}\left(\textrm{j}\left(\omega t-kz\right)\right)$, where $k=n\omega/c$ denotes the wavenumber, $c$ is the speed of light, and $n$ is the refractive index of the respective medium. The signal comb consists of discrete tones at frequencies $\omega_{\rm{S},\mu} = \omega_{\rm{S},0} + \mu \cdot \omega_{\rm{S,r}}$. For the first part of the signal comb, these tones accumulate phase shifts $\Psi_{\rm{meas},\mu}$ when propagating back and forth over the measurement distance $d$ to the target. The other part of the signal comb is directly guided to the reference detector (Reference PD), thereby accumulating phase shifts of $\Psi_{\rm{ref},\mu}$,
\begin{align}
\Psi_{\rm{meas},\mu}&=-\frac{\omega_{\rm{S},\mu}}{ c}\left(L_{\rm{meas}}+2d\right),\nonumber\\
\Psi_{\rm{ref},\mu}&=-\frac{\omega_{\rm{S},\mu}}{ c}L_{\rm{ref}}. \label{Eq.1}
\end{align}
\noindent In this relation, $L_{\rm{meas}}$ and $L_{\rm{ref}}$ refer to the optical path lengths that the two parts of the signal comb propagate on the chips from the common comb generator to the respective photodetector, and $d$ is the single-pass free-space distance towards the target. The two parts of the signal comb are superimposed with portions of the LO comb on the respective photodetector, thus leading to a multitude of sinusoidal signals in the corresponding baseband photocurrents, see Supplementary Information. In essence, the phases of these sinusoidals reveal the differences $\Delta\Psi_{\mu}(d) = \Psi_{\rm{ref,}\mu}-\Psi_{\rm{meas},\mu} $ of the phases $\Psi_{\rm{meas},\mu}$ and $\Psi_{\rm{ref},\mu}$ that the respective signal comb tones have accumulated along the measurement and the reference path. To eliminate the internal optical path lengths $L_{\rm{meas}}$ and $L_{\rm{ref}}$, a calibration measurement at a known distance $d_0$ is performed. The target distance $\left(d-d_0\right)$ can then be extracted by estimating the slope of the phase differences $\Delta\Phi_{\mu}=\Delta\Psi_{\mu}(d)-\Delta\Psi_{\mu}(d_0)$ as a function of frequency index $\mu$, % =(L_{\rm{tar}}-L_{\rm{ref}})/2$ 
\begin{align}
\Delta\Phi_{\mu}&=\Delta\Phi_{0}+\mu\frac{ \omega_{\rm{S,r}}}{ c}\times 2\left(d-d_{0}\right),\nonumber\\
\Delta\Phi_{0}&=\frac{\omega_{\rm{S,0}}}{ c}\times 2\left(d-d_{0}\right). \label{Eq.2}
\end{align}
\noindent In this relation, $\Delta\Phi_{0}$ denotes a phase offset that is independent of $\mu$. The slope of $\Delta\Phi_{\mu}$ with respect to $\mu$ is extracted from the measured phases by means of a linear fit. Note that this technique allows to directly check the validity of a certain distance measurement by using the fit error as a quality criterion -- unreliable raw data leads to large fit errors, based on which erroneous points can be discarded, see Supplementary Information.

The unique advantages of DKS combs for high-speed high-precision sampling can be understood by analyzing the fundamental limitations of measurement accuracy and acquisition speed. Note that, for high-speed sampling, it is important to keep the number $N$ of baseband beat notes as small as possible: For $N$ beat notes that are equally distributed over a given acquisition bandwidth $f_{\rm{ADC}}$ of the ADC, the frequency spacing $\Delta f_{\rm{r}} = \Delta\omega_{\rm{r}} / 2\pi$ of the beat notes is at most $f_{\rm{ADC}}/N$. To spectrally resolve these beat notes by a Fourier transformation, a minimum observation time of $T_{\rm{min}}=1 /\Delta f_{\rm{r}}\geq N/f_{\rm{ADC}}$ is required. This leads to a maximum distance acquisition rate of $T_{\rm{min}}^{-1} = \Delta f_{\rm{r}} \leq f_{\rm{ADC}}/N$.

On the other hand, the number $N$ of optical tones used for distance measurement will also influence the accuracy with which we can estimate the slope of the phase differences $\Delta\Phi_{\mu}$ vs. $\mu$ from the noisy measurement data. Using basic relations of linear regression analysis, the standard deviation of the measured distance can be estimated from the overall optical bandwidth $\Omega_{\rm{S}}=N\omega_{\rm{S,r}}$ of the comb, the standard deviation $\sigma_{\phi}$ of the individual phase measurements, and the number $N$ of optical tones, see Supplementary Information for details,
\begin{equation}
\sigma_{d} = \sqrt{\frac{3}{N}}\frac{c}{\Omega_{\rm{S}}} \sigma_{\phi}. \label{Eq.3} 
\end{equation}
\noindent For a fixed number $N$ of optical lines, the only option that remains for improving the measurement accuracy is to increase the overall optical bandwidth $\Omega_{\rm{S}}=N\omega_{\rm{S,r}}$ of the comb, which requires a comb source that provides a large free spectral range. DKS combs stand out due to a unique combination of large overall optical bandwidth and large FSR. They hence feature comparatively few optical lines and are thus perfectly suited for simultaneously achieving high sampling and high measurement accuracy. This is demonstrated in the following section by using a simplified model system that combines a pair of DKS comb sources with fiber-optic components and a high-speed oscilloscope for data acquisition and offline processing.

\begin{figure*}
\includegraphics[width = \textwidth]{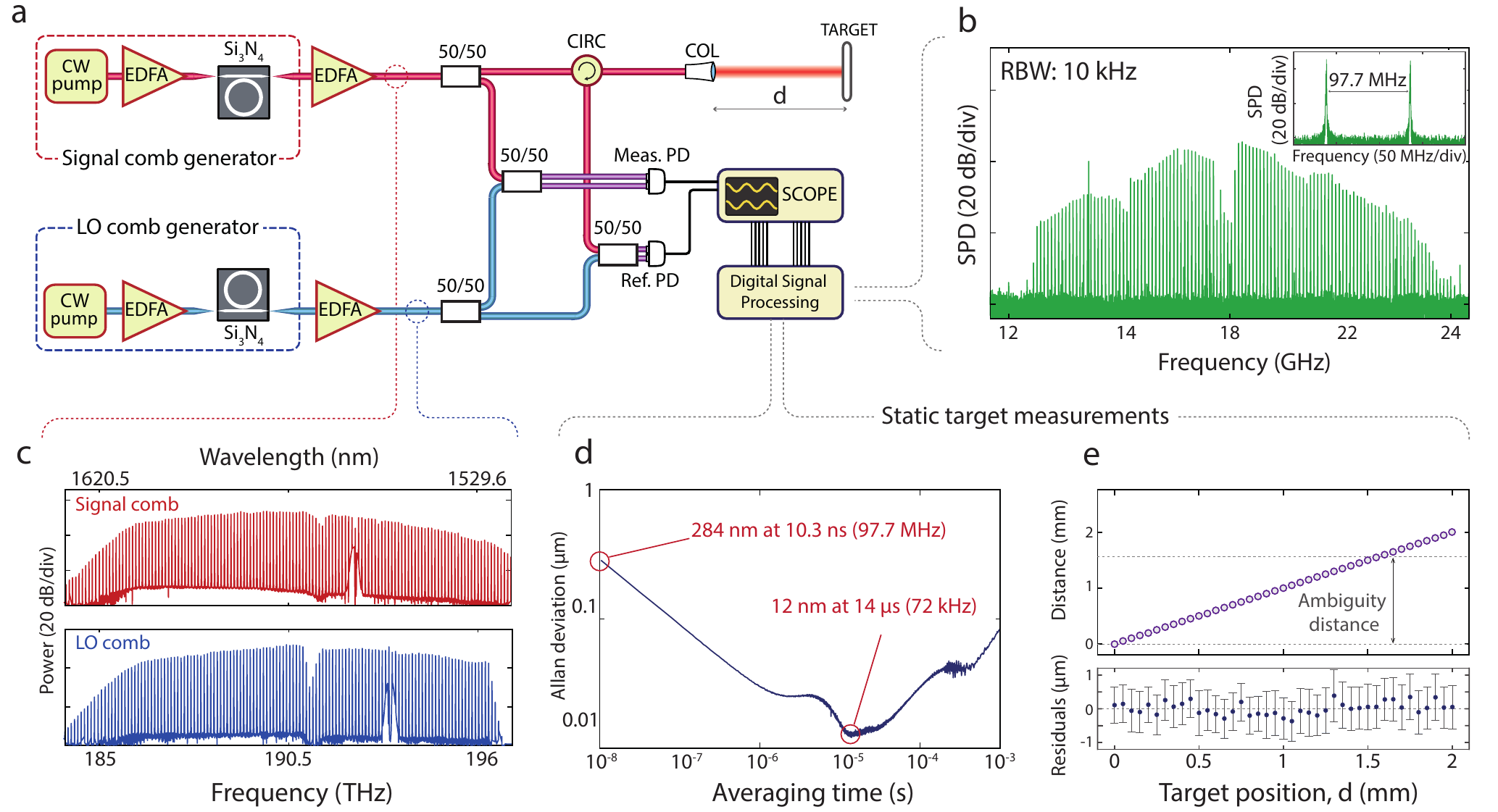}
\protect\caption{\textbf{\boldmath Experimental demonstration and performance characterization.}
\textbf{(a)} Experimental setup. DKS combs are generated by a pair of $\rm Si_{3}N_{4}$ microresonators, which are pumped by two CW lasers and erbium-doped fiber amplifiers (EDFA) -- details on the comb generation scheme can be found in the Supplementary Information. The combs are detuned in line spacing by $\left|\omega_{\rm{LO,r}} - \omega_{\rm{S,r}}\right|/2\pi \approx$ 98\,MHz as well as in center frequency by $\Delta\omega_{0}/2\pi = \left|\omega_{\rm{LO,0}} - \omega_{\rm{S,0}}\right|/2\pi \approx$ 18\,GHz. After suppressing residual pump light by a fiber Bragg gratings (not shown), the combs are amplified by another pair of EDFA. The signal comb (red) is split and one part is routed to the target and back to a balanced measurement PD (Meas.\,PD) by optical fibers, an optical circulator (CIRC) and a collimator (COL), while the other part is directly sent to the balanced reference detector (Ref.\,PD). Similarly, the LO comb is split in two portions, which are routed to the measurement PD and the reference PD for multi-heterodyne detection. The resulting baseband beat signals are recorded by a 32\,GHz real-time sampling oscilloscope. As a target, we use a silver mirror that can be positioned with an accuracy of better than 50\,nm using a feedback-stabilized stage. 
\textbf{(b)} Numerically calculated Fourier transform of a recorded time-domain signal. The baseband spectrum consists of discrete spectra centered around $\Delta\omega_{0}/2\pi = 18\,\textrm{GHz}$ with spacings of $\left|\omega_{\rm{LO,r}} - \omega_{\rm{S,r}}\right|/2\pi \approx$ 98 MHz.
\textbf{(c)} Optical spectra of the signal and the LO comb after amplification (red / blue). Both combs cover a range of approximately 11\,THz, limited by the gain bandwidth of the respective EDFA. The spectral ranges of the signal and LO combs amount to $\omega_{\rm{S,r}}/2\pi = $ 95.646\,GHz and $\omega_{\rm{LO,r}}/2\pi = $ 95.549\,GHz, respectively. Both spectra feature a slight dip near 191\,THz, which is caused by internal multiplexing and demultiplexing filters of the EDFA. The increase of the noise background around 192.50\,THz (193.46\,THz) in the signal (LO) comb is caused by residual amplified spontaneous emission (ASE) originating from the pump EDFA \cite{Marin-Palomo2016}.
\textbf{(d)} Allan deviation of measured distances as a function of averaging time. The highest acquisition speed is limited by the spacing of the baseband beat notes in the photocurrent and amounts to approximately 98\,MHz (acquisition time of 10.3\,ns). At this speed, an Allan deviation of 284\,nm is achieved, that decreases to a record-low value of 12\,nm at an averaging time of 14\,$\mu$s. The increase towards longer averaging times is attributed to drifts in the fiber-optic setup, see Supplementary Information.
\textbf{(e)} Top: Scan of measured position vs. set position in steps of 50\,$\mu$m over the full ambiguity distance (marked by dashed lines). The points outside the ambiguity interval are manually unwrapped. Bottom: Residual deviations ("residuals") between measured and set positions. The residuals are of the same order of magnitude as the 50\,nm positioning accuracy of the positioning stage. Error bars indicate the standard deviation obtained for each position of the mirror, see Supplementary Information for details of the distance sweep experiment. Importantly, the residuals do not show any cyclic error.
\label{fig_2}}
\end{figure*} 

%%%%%%%%%%%%%%%%%%%%%%%%%%%%%%%%%%%%%%%%%%%%%%%%
%%%%%%%%%%%%%%%%%%%%%%%%%%%%%%%%%%%%%%%%%%%%%%%%
\section*{Results}
%%%%%%%%%%%%%%%%%%%%%%%%%%%%%%%%%%%%%%%%%%%%%%%%
%%%%%%%%%%%%%%%%%%%%%%%%%%%%%%%%%%%%%%%%%%%%%%%%

In the following, we demonstrate the basic viability of the envisaged chip-scale LIDAR engine introduced in Fig.\,\ref{fig_1}, which critically relies on integrated DKS comb generators as core components. Note that all other integrated components of the LIDAR engine rely on the standard device portfolio of photonic integration and have been established in prior work \cite{Roelkens2010,Billah2017,Weimann2014}.
The experimental setup for our proof-of-concept demonstration is depicted in Fig.\,\ref{fig_2}a. For DKS comb generation, we use a pair of CW-pumped $\rm Si_{3}N_{4}$-microring resonators on separate chips. The devices are fabricated using the photonic Damascene process \cite{Pfeiffer2015Damascene} with waveguide cross sections of $1.65\,\mu \textrm{m} \times 0.8\,\mu\textrm{m}$. For multi-heterodyne detection, the two resonators have slightly different free spectral ranges of $\omega_{\rm{S,r}}/2\pi = $ 95.646\,GHz and $\omega_{\rm{LO,r}}/2\pi = $ 95.549\,GHz respectively. To demonstrate that our concept does not require phase locking of the DKS combs, we used a pair of free-running pump lasers, even though a single pump laser could have been used as well \cite{Joshi2017,yu2016DKSdualcombMIR}. The pump light is amplified in erbium-doped fiber amplifiers (EDFA) and then coupled to the microresonator chips. DKS comb generation is achieved by sweeping the pump laser frequency from the effectively blue-detuned to a defined point in the effectively red-detuned regime of a selected cavity resonance, where the microresonator system supports the soliton formation\cite{Karpov2016}. By further applying the backward tuning technique \cite{Karpov2016}, a single-soliton state corresponding to an optical frequency comb with spectrally smooth $\rm sech^2$-shape envelope is deterministically achieved. A fiber Bragg grating (not shown) is used to suppress remaining pump light. A more detailed description of the experimental setup and of the microresonator devices can be found in the Supplementary Information.

The resulting combs are amplified in C+L band EDFA to improve phase extraction of the individual beat notes. The spectra of the amplified combs are shown in Fig.\,\ref{fig_2}c for the signal (red) and the LO (blue) comb. The gain bandwidth of the EDFA limits the number of usable lines to about 115, which is sufficient for our experiments. 

For distance measurement, the signal comb (red) is split by a fiber-based 50:50 coupler, and one part is routed to the target and back to a balanced measurement PD (meas.\,PD), while the other part is directly sent to the balanced reference detector (ref. PD), see Fig.\,\ref{fig_2}a. Measurement and reference PD feature bandwidths of 43\,GHz. Note that, in contrast to Fig.\,\ref{fig_1}a and Fig.\,\ref{fig_1}c, we use additionally a circulator in conjunction with a directional coupler for splitting forward and backward-propagating light in the measurement path. Similarly, the LO comb is split in two portions, which are routed to the measurement PD and the reference PD for multi-heterodyne detection. The resulting baseband beat signals are recorded by a 32\,GHz real-time sampling oscilloscope. Data processing and evaluation is performed offline. 

Figure\,\ref{fig_2}b shows a numerically calculated Fourier transform of a recorded time-domain signal that reveals a multitude of discrete beat notes between the signal and LO comb lines. The spacing of the beat notes is given by the difference of the line spacing of the LO and the signal comb and amounts to $\Delta f_{\rm{r}} = \Delta\omega_{\rm{r}}/2\pi$ = 97.7\,MHz, thereby dictating a minimum possible acquisition time of $T_{\rm{min}} = 1/\Delta f_{\rm{r}} = 10.3$\,ns and a maximum possible distance acquisition rate of 97.7\,MHz. This is the highest value demonstrated so far exceeding the fastest previous demonstration \cite{Ataie2013} by more than an order of magnitude. 

For a thorough stability and precision analysis of our dual-comb scheme, we evaluate the Allan deviation \cite{Allan.1966} over a 10.3\,ms long measurement that is composed of a series of $10^6$ individual data points taken from a static mirror at an acquisition time of 10.3\,ns per point. The results are plotted in Fig.\,\ref{fig_2}d. At an averaging time of 10.3\,ns, an Allan deviation of 284\,nm is obtained, that decreases to 12\,nm for an averaging time of 14\,$\mu$s. To the best of our knowledge, this is the lowest Allan deviation so far demonstrated by comb-based synthetic-wavelength interferometry. The initial linear decrease of the Allan deviation implies dominating high-frequency noise, which is attributed to an amplified spontaneous emission (ASE) background originating from the deployed EDFA. This ASE noise impacts the phase estimation of the individual beat notes and therefore the extracted distance. For longer averaging times, the Allan deviation increases, which we attribute to thermal drift of the fibers and to mechanical vibrations.
 
Besides the Allan deviation of a distance measurement to a static target, we estimated the accuracy of our technique for measuring variable distances to a target that is moved over a full ambiguity distance, see Fig\,\ref{fig_2}d. In this experiment, the target mirror is stepped in increments of 50\,$\mu$m using a high-precision translation stage with an accuracy of better than 50\,nm. To eliminate the impact of fiber drift, the distance measurement is continuously switched between the movable target mirror and a static calibration mirror in quick succession, taking between 6\,500 and 9\,500 measurements on each mirror, see Supplementary Information for details. To minimize the impact of high-frequency noise, an averaging time of 100\,$\mu$s is chosen. In the upper part of Fig.\,\ref{fig_2}e, the measured distance is plotted as a function of the distance set by the translation state. Measured distances exceeding the ambiguity interval of 1.51\,mm are unwrapped manually. The bottom part of Fig.\,\ref{fig_2}e shows the deviations of the measured from the set positions along with the respective standard deviations indicated as error bars. Importantly, no cyclic error is observed throughout the ambiguity interval. We determine the accuracy of our measurement to 188\,nm, defined as the standard deviation of the residuals. These residuals are of the same order of magnitude as the 50\,nm positioning accuracy of the positioning stage specified by the manufacturer \cite{pipositioner}. In this measurement, the refractive index of air is considered according to Ciddors formula for ambient lab conditions~\cite{Ciddor.1996}. 

%CK: Ab hier

\begin{figure*}
\includegraphics[width = \textwidth]{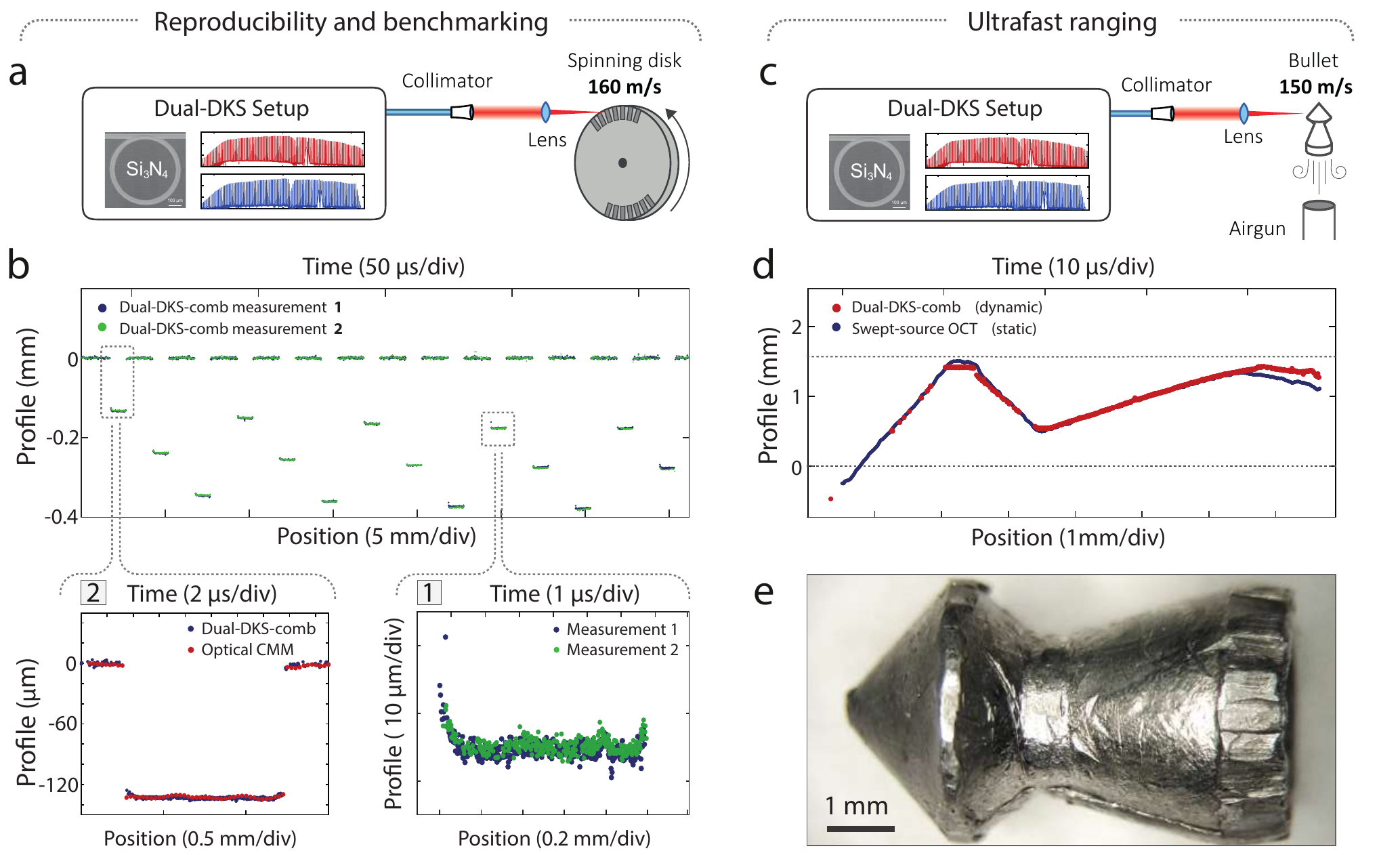}
\protect\caption{\textbf{\boldmath Reproducibility, benchmarking and ultrafast ranging demonstration.}
\textbf{(a)} Setup for reproducibility and benchmarking experiments: The measurement beam is focused on the surface of a spinning disk having grooves of different depths on its surface. The disk rotates at a frequency of about $600\,\rm{Hz}$, resulting in an edge velocity of 160 m/s. Unreliable measurement points close to the edges of the grooves lead to a large fit error of the linear phase characteristic according to Eq.\,\ref{Eq.2} and can thus be discarded.
\textbf{(b)} Measured surface profile of the disk as a function of position (bottom) and time (top) for two independent measurements at the same position.
\textbf{Inset\,1:} Reproducibility demonstration by detailed comparison of the two independent measurements plotted in (b). The results exhibit good agreement regarding both macroscopic features such as the groove depth and width as well as microscopic features such as surface texture and a decrease of depth towards the edge of the groove. Deviations are attributed to the fact that the two measurements have been taken independently and might hence not have sampled the exactly same line across the groove.
\textbf{Inset\,2:} Benchmarking of the high-speed dual-DKS-comb measurement to the results obtained from an industrial optical coordinate-measuring machine (CMM). Both profiles are in good agreement. Deviations are attributed to slightly different measurement positions within the analyzed groove.
\textbf{(c)} Setup for ultrafast ranging experiment. The measurement beam is focused into free space. An air-gun bullet is then fired through the focal spot of the beam and the profile is recorded. The bullet travels at a speed of $\sim$150\,m/s (Mach 0.47).
\textbf{(d)} Measured profile of the projectile obtained from in-flight dual-DKS-comb measurement (red), along with a swept-source OCT profile scan that was recorded on the static projectile after recovery from the backstop. For better comparison, the two profiles were rotated and an actual speed of the bullet of 149\,m/s was estimated for best agreement of the two profiles. Both curves show good agreement, demonstrating the ability of the dual-DKS-comb technique to obtain reliable results even for ultrafast sampling on rapidly moving targets with naturally scattering surfaces. Missing data points in the dual-DKS-comb measurement at the tip of the projectile are caused by low power levels of the back-coupled signal, which is inevitable for such steep surfaces in combination with the limited numerical aperture of the lens used for focusing the beam.  Deviations towards the back end of the bullet are attributed to the fact that the strongly corrugated surface in this area has been sampled at two different positions.
\textbf{(e)} Image of the projectile after being fired. Deformations during the shot have lead to a strong corrugation of the bullet towards its back.}
\label{fig_3}
\end{figure*} 

\noindent To validate the reproducibility of our system and to benchmark the results with respect to existing techniques, we measure the profile of a quickly rotating disk having grooves on its surface, see Fig.\,\ref{fig_3}a. In this experiment, the measurement beam is focused to the surface near the edge of the disk, which rotates at an frequency of about 600\,Hz, resulting in an edge velocity of 160\,m/s. The distance acquisition rate amounts to 97.1\,MHz, limited by the spectral spacing of $\Delta\omega_{\rm{r}}/2\pi$ of the beat notes in the baseband photocurrent, but not by the acquisition speed of our oscilloscopes. The resulting profiles are shown in Fig.\,\ref{fig_3}b for two measurements, which were taken independently from one another. Measurement points close to the edges of the grooves may suffer from strong scattering and low power levels, which lead to unreliable distance information. Using the fit error of the linear phase characteristic according to Eq.\,\ref{Eq.2} as a quality criterion, our technique allows to identify such bad measurement points and to automatically discard them from the data, see Supplementary Information for details. The raw data of both measurements was further subject to vibrations of the disk arising from the driving engine. These vibrations have been removed by fitting a polynomial to the top surface of the disk and by using it for correction of the overall measurement data. In a first experiment, we analyze the reproducibility of the technique by a detailed comparison of the results obtained from the two measurements, see Fig.\,\ref{fig_3}b, Inset\,1. The measured profiles exhibit good agreement regarding both macroscopic features such as the groove depth and width as well as microscopic features  such  as  surface  texture  and  a  decrease  of  depth  towards  the  edge  of  the  groove. Deviations are attributed to the fact that the two measurements have been taken independently and might hence not have sampled the exactly same line across the groove. 

In addition, we benchmark our technique by comparing the obtained profile of a single groove with a profile obtained from an industrial optical coordinate-measuring machine (CMM, Werth VideoCheck HA), Fig.\,\ref{fig_3}b, Inset\,2. Both profiles are in good agreement, with some minor deviations that we attribute to slightly different measurement positions within the analyzed groove.

In a final experiment, we demonstrate ultrafast ranging by measuring the profile of a flying air-gun bullet that is shot through the focus of the measurement beam, see Fig.\,\ref{fig_3}c. The projectile moves at a speed of 150\,m/s, i.e. Mach 0.47, which, together with the acquisition rate of 97.7\, MHz, results in a lateral distance of 1.5\,$\mu$m between neighbouring sampling points on the surface of the bullet. The measured profile is depicted in red in Fig.\,\ref{fig_3}d along with a reference measurement of the profile obtained from the static bullet using a swept-source OCT system (dark blue). Both curves clearly coincide and reproduce the shape of the fired projectile that can be seen in Fig.\,\ref{fig_3}e. Missing data points in the dual-DKS-comb measurement at the tip of the projectile are caused by low power levels of the back-coupled signal, which is inevitable for such steep surfaces in combination with the limited numerical aperture of the lens used for focussing the beam. These measurement points have been discarded from the data based on a large fit error of the linear phase characteristic according to Eq.\,\ref{Eq.2}. An image of the projectile after recovery from the backstop exhibits a strong corrugation of the bullet towards its back, Fig.\,\ref{fig_3}e. This leads to deviations of the measured profiles in Fig.\,\ref{fig_3}d towards the right-hand side, since the strongly corrugated surface of the projectile in this area has very likely been sampled at two different positions. These experiments clearly demonstrate the viability of the dual-DKS-comb approach and its extraordinary performance advantages for ultrafast high-precision sampling. Thanks to the high sampling rate, the technique would allow to track continuous movements of objects at any practical speed, with an ambiguity limit 144\,000\,m/s. The ambiguity distance of the dual-DKS-comb approach can be greatly increased by combination with a comparatively simple  low-accuracy time-of-flight system.

\section*{Summary}
We have shown ultrafast high-precision optical ranging using a pair of microresonator-based dissipative Kerr soliton frequency combs. DKS combs offer a unique combination of large optical bandwidth and large free spectral range, thereby enabling the fastest ranging experiment to date. We achieve distance acquisition rates of 97.7 MHz while maintaining sub-$\mu$m precision on a macroscopic scale, thereby outperforming the fastest previous demonstrations by more than an order of magnitude. We investigate the reproducibility of our system, benchmark it with respect to an industrial coordinate measuring machine (CMM), and finally demonstrate its performance by sampling the naturally scattering surface of air-gun bullets on the fly. Our results may impact both scientific and industrial applications that require fast and precise contact-less distance measurements. The scheme is fully amenable to photonic integration, thereby offering a promising route towards cost-efficient mass-production of compact LIDAR engines with ultra-high sampling rates.  
\section*{Acknowledgments}
\label{Acknowledgements}
This work was supported by the Deutsche Forschungsgemeinschaft (DFG) through the Collaborative Research Center 'Wave Phenomena: Analysis and Numerics' (CRC 1173), project B3 'Frequency combs', by the European Research Council (ERC Starting Grant 'EnTeraPIC', number 280145), by the EU project BigPipes, by the Alfried Krupp von Bohlen und Halbach Foundation, by the Karlsruhe School of Optics and Photonics (KSOP), and by the Helmholtz International Research School for Teratronics (HIRST). D.G. is supported by the Horizon 2020 Research and Innovation Program under the Marie Sklodowska-Curie Grant Agreement no. 642890 (TheLink). P.M.-P. is supported by the Erasmus Mundus doctorate programme Europhotonics (grant number 159224-1-2009-1-FR-ERA MUNDUS-EMJD). 
$\rm Si_3N_4$ devices were fabricated and grown in the Center of MicroNanoTechnology (CMi) at EPFL. EPFL acknowledges support by the Air Force Office of Scientific Research, Air Force Material Command, USAF, number FA9550-15-1-0099.
This work is supported by the European Space Technology Centre with ESA Contract No.:4000116145/16/NL/MH/GM.
M.K. acknowledges funding support from Marie Curie FP7 ITN FACT.

\subsection*{Note added}
During the preparation of this manuscript, dual-soliton-comb distance measurements were also reported in silica microdisk resonators \cite{suh2017VahalaLIDAR}.

% Bib file
%\bibliography{DualSolitonDistanceLibrary}

\begin{thebibliography}{35}%
\makeatletter
\providecommand \@ifxundefined [1]{%
 \@ifx{#1\undefined}
}%
\providecommand \@ifnum [1]{%
 \ifnum #1\expandafter \@firstoftwo
 \else \expandafter \@secondoftwo
 \fi
}%
\providecommand \@ifx [1]{%
 \ifx #1\expandafter \@firstoftwo
 \else \expandafter \@secondoftwo
 \fi
}%
\providecommand \natexlab [1]{#1}%
\providecommand \enquote  [1]{``#1''}%
\providecommand \bibnamefont  [1]{#1}%
\providecommand \bibfnamefont [1]{#1}%
\providecommand \citenamefont [1]{#1}%
\providecommand \href@noop [0]{\@secondoftwo}%
\providecommand \href [0]{\begingroup \@sanitize@url \@href}%
\providecommand \@href[1]{\@@startlink{#1}\@@href}%
\providecommand \@@href[1]{\endgroup#1\@@endlink}%
\providecommand \@sanitize@url [0]{\catcode `\\12\catcode `\$12\catcode
  `\&12\catcode `\#12\catcode `\^12\catcode `\_12\catcode `\%12\relax}%
\providecommand \@@startlink[1]{}%
\providecommand \@@endlink[0]{}%
\providecommand \url  [0]{\begingroup\@sanitize@url \@url }%
\providecommand \@url [1]{\endgroup\@href {#1}{\urlprefix }}%
\providecommand \urlprefix  [0]{URL }%
\providecommand \Eprint [0]{\href }%
\providecommand \doibase [0]{http://dx.doi.org/}%
\providecommand \selectlanguage [0]{\@gobble}%
\providecommand \bibinfo  [0]{\@secondoftwo}%
\providecommand \bibfield  [0]{\@secondoftwo}%
\providecommand \translation [1]{[#1]}%
\providecommand \BibitemOpen [0]{}%
\providecommand \bibitemStop [0]{}%
\providecommand \bibitemNoStop [0]{.\EOS\space}%
\providecommand \EOS [0]{\spacefactor3000\relax}%
\providecommand \BibitemShut  [1]{\csname bibitem#1\endcsname}%
\let\auto@bib@innerbib\@empty
%</preamble>
\bibitem [{\citenamefont {Amann}\ \emph {et~al.}(2001)\citenamefont {Amann},
  \citenamefont {Bosch}, \citenamefont {Lescure}, \citenamefont {Myllyla},\
  and\ \citenamefont {Rioux}}]{amann2001laserRangingReview}%
  \BibitemOpen
  \bibfield  {author} {\bibinfo {author} {\bibfnamefont {M.-C.}\ \bibnamefont
  {Amann}}, \bibinfo {author} {\bibfnamefont {T.}~\bibnamefont {Bosch}},
  \bibinfo {author} {\bibfnamefont {M.}~\bibnamefont {Lescure}}, \bibinfo
  {author} {\bibfnamefont {R.}~\bibnamefont {Myllyla}}, \ and\ \bibinfo
  {author} {\bibfnamefont {M.}~\bibnamefont {Rioux}},\ }\href@noop {}
  {\bibfield  {journal} {\bibinfo  {journal} {Optical Engineering}\ }\textbf
  {\bibinfo {volume} {40}},\ \bibinfo {pages} {10} (\bibinfo {year}
  {2001})}\BibitemShut {NoStop}%
\bibitem [{\citenamefont {Berkovic}\ and\ \citenamefont
  {Shafir}(2012)}]{Berkovic2012}%
  \BibitemOpen
  \bibfield  {author} {\bibinfo {author} {\bibfnamefont {G.}~\bibnamefont
  {Berkovic}}\ and\ \bibinfo {author} {\bibfnamefont {E.}~\bibnamefont
  {Shafir}},\ }\href {\doibase 10.1364/AOP.4.000441} {\bibfield  {journal}
  {\bibinfo  {journal} {Advances in Optics and Photonics}\ }\textbf {\bibinfo
  {volume} {4}},\ \bibinfo {pages} {441} (\bibinfo {year} {2012})}\BibitemShut
  {NoStop}%
\bibitem [{\citenamefont {Park}\ and\ \citenamefont
  {Kim}(1994)}]{park1994opticalMeasSpindel}%
  \BibitemOpen
  \bibfield  {author} {\bibinfo {author} {\bibfnamefont {Y.-C.}\ \bibnamefont
  {Park}}\ and\ \bibinfo {author} {\bibfnamefont {S.-W.}\ \bibnamefont {Kim}},\
  }\href@noop {} {\bibfield  {journal} {\bibinfo  {journal} {International
  Journal of Machine Tools and Manufacture}\ }\textbf {\bibinfo {volume}
  {34}},\ \bibinfo {pages} {1019} (\bibinfo {year} {1994})}\BibitemShut
  {NoStop}%
\bibitem [{\citenamefont {Levinson}\ \emph {et~al.}(2011)\citenamefont
  {Levinson}, \citenamefont {Askeland}, \citenamefont {Becker}, \citenamefont
  {Dolson}, \citenamefont {Held}, \citenamefont {Kammel}, \citenamefont
  {Kolter}, \citenamefont {Langer}, \citenamefont {Pink}, \citenamefont {Pratt}
  \emph {et~al.}}]{levinson2011towardsAutonomousDriving}%
  \BibitemOpen
  \bibfield  {author} {\bibinfo {author} {\bibfnamefont {J.}~\bibnamefont
  {Levinson}}, \bibinfo {author} {\bibfnamefont {J.}~\bibnamefont {Askeland}},
  \bibinfo {author} {\bibfnamefont {J.}~\bibnamefont {Becker}}, \bibinfo
  {author} {\bibfnamefont {J.}~\bibnamefont {Dolson}}, \bibinfo {author}
  {\bibfnamefont {D.}~\bibnamefont {Held}}, \bibinfo {author} {\bibfnamefont
  {S.}~\bibnamefont {Kammel}}, \bibinfo {author} {\bibfnamefont {J.~Z.}\
  \bibnamefont {Kolter}}, \bibinfo {author} {\bibfnamefont {D.}~\bibnamefont
  {Langer}}, \bibinfo {author} {\bibfnamefont {O.}~\bibnamefont {Pink}},
  \bibinfo {author} {\bibfnamefont {V.}~\bibnamefont {Pratt}},  \emph
  {et~al.},\ }in\ \href@noop {} {\emph {\bibinfo {booktitle} {Intelligent
  Vehicles Symposium (IV), 2011 IEEE}}}\ (\bibinfo {organization} {IEEE},\
  \bibinfo {year} {2011})\ pp.\ \bibinfo {pages} {163--168}\BibitemShut
  {NoStop}%
\bibitem [{\citenamefont {Sassen}\ \emph {et~al.}(2008)\citenamefont {Sassen},
  \citenamefont {Wang},\ and\ \citenamefont {Liu}}]{sassen2008clouds2Lidars}%
  \BibitemOpen
  \bibfield  {author} {\bibinfo {author} {\bibfnamefont {K.}~\bibnamefont
  {Sassen}}, \bibinfo {author} {\bibfnamefont {Z.}~\bibnamefont {Wang}}, \ and\
  \bibinfo {author} {\bibfnamefont {D.}~\bibnamefont {Liu}},\ }\href@noop {}
  {\bibfield  {journal} {\bibinfo  {journal} {Journal of Geophysical Research:
  Atmospheres}\ }\textbf {\bibinfo {volume} {113}} (\bibinfo {year}
  {2008})}\BibitemShut {NoStop}%
\bibitem [{\citenamefont {Li}\ \emph {et~al.}(2014)\citenamefont {Li},
  \citenamefont {Liu}, \citenamefont {Zhang},\ and\ \citenamefont
  {Hang}}]{li2014lidarDrone}%
  \BibitemOpen
  \bibfield  {author} {\bibinfo {author} {\bibfnamefont {R.}~\bibnamefont
  {Li}}, \bibinfo {author} {\bibfnamefont {J.}~\bibnamefont {Liu}}, \bibinfo
  {author} {\bibfnamefont {L.}~\bibnamefont {Zhang}}, \ and\ \bibinfo {author}
  {\bibfnamefont {Y.}~\bibnamefont {Hang}},\ }in\ \href@noop {} {\emph
  {\bibinfo {booktitle} {Inertial Sensors and Systems Symposium (ISS), 2014
  DGON}}}\ (\bibinfo {organization} {IEEE},\ \bibinfo {year} {2014})\ pp.\
  \bibinfo {pages} {1--15}\BibitemShut {NoStop}%
\bibitem [{\citenamefont {Udem}\ \emph {et~al.}(2002)\citenamefont {Udem},
  \citenamefont {Holzwarth},\ and\ \citenamefont {H{\"{a}}nsch}}]{Udem2002}%
  \BibitemOpen
  \bibfield  {author} {\bibinfo {author} {\bibfnamefont {T.}~\bibnamefont
  {Udem}}, \bibinfo {author} {\bibfnamefont {R.}~\bibnamefont {Holzwarth}}, \
  and\ \bibinfo {author} {\bibfnamefont {T.~W.}\ \bibnamefont {H{\"{a}}nsch}},\
  }\href {\doibase 10.1038/416233a} {\bibfield  {journal} {\bibinfo  {journal}
  {Nature}\ }\textbf {\bibinfo {volume} {416}},\ \bibinfo {pages} {233}
  (\bibinfo {year} {2002})}\BibitemShut {NoStop}%
\bibitem [{\citenamefont {Minoshima}\ and\ \citenamefont
  {Matsumoto}(2000)}]{minoshima2000combLIDAR}%
  \BibitemOpen
  \bibfield  {author} {\bibinfo {author} {\bibfnamefont {K.}~\bibnamefont
  {Minoshima}}\ and\ \bibinfo {author} {\bibfnamefont {H.}~\bibnamefont
  {Matsumoto}},\ }\href@noop {} {\bibfield  {journal} {\bibinfo  {journal}
  {Applied Optics}\ }\textbf {\bibinfo {volume} {39}},\ \bibinfo {pages} {5512}
  (\bibinfo {year} {2000})}\BibitemShut {NoStop}%
\bibitem [{\citenamefont {Schuhler}\ \emph {et~al.}(2006)\citenamefont
  {Schuhler}, \citenamefont {Salvad{\'e}}, \citenamefont {L{\'e}v{\^e}que},
  \citenamefont {D{\"a}ndliker},\ and\ \citenamefont
  {Holzwarth}}]{schuhler2006comb2wavelength}%
  \BibitemOpen
  \bibfield  {author} {\bibinfo {author} {\bibfnamefont {N.}~\bibnamefont
  {Schuhler}}, \bibinfo {author} {\bibfnamefont {Y.}~\bibnamefont
  {Salvad{\'e}}}, \bibinfo {author} {\bibfnamefont {S.}~\bibnamefont
  {L{\'e}v{\^e}que}}, \bibinfo {author} {\bibfnamefont {R.}~\bibnamefont
  {D{\"a}ndliker}}, \ and\ \bibinfo {author} {\bibfnamefont {R.}~\bibnamefont
  {Holzwarth}},\ }\href@noop {} {\bibfield  {journal} {\bibinfo  {journal}
  {Optics Letters}\ }\textbf {\bibinfo {volume} {31}},\ \bibinfo {pages} {3101}
  (\bibinfo {year} {2006})}\BibitemShut {NoStop}%
\bibitem [{\citenamefont {Coddington}\ \emph {et~al.}(2009)\citenamefont
  {Coddington}, \citenamefont {Swann}, \citenamefont {Nenadovic},\ and\
  \citenamefont {Newbury}}]{Coddington2009}%
  \BibitemOpen
  \bibfield  {author} {\bibinfo {author} {\bibfnamefont {I.}~\bibnamefont
  {Coddington}}, \bibinfo {author} {\bibfnamefont {W.~C.}\ \bibnamefont
  {Swann}}, \bibinfo {author} {\bibfnamefont {L.}~\bibnamefont {Nenadovic}}, \
  and\ \bibinfo {author} {\bibfnamefont {N.~R.}\ \bibnamefont {Newbury}},\
  }\href {\doibase 10.1038/nphoton.2009.94} {\bibfield  {journal} {\bibinfo
  {journal} {Nature Photonics}\ }\textbf {\bibinfo {volume} {3}},\ \bibinfo
  {pages} {351} (\bibinfo {year} {2009})}\BibitemShut {NoStop}%
\bibitem [{\citenamefont {Jang}\ \emph {et~al.}(2016)\citenamefont {Jang},
  \citenamefont {Wang}, \citenamefont {Hyun}, \citenamefont {Kang},
  \citenamefont {Chun}, \citenamefont {Kim},\ and\ \citenamefont
  {Kim}}]{Jang.2016}%
  \BibitemOpen
  \bibfield  {author} {\bibinfo {author} {\bibfnamefont {Y.-S.}\ \bibnamefont
  {Jang}}, \bibinfo {author} {\bibfnamefont {G.}~\bibnamefont {Wang}}, \bibinfo
  {author} {\bibfnamefont {S.}~\bibnamefont {Hyun}}, \bibinfo {author}
  {\bibfnamefont {H.~J.}\ \bibnamefont {Kang}}, \bibinfo {author}
  {\bibfnamefont {B.~J.}\ \bibnamefont {Chun}}, \bibinfo {author}
  {\bibfnamefont {Y.-J.}\ \bibnamefont {Kim}}, \ and\ \bibinfo {author}
  {\bibfnamefont {S.-W.}\ \bibnamefont {Kim}},\ }\href {\doibase
  10.1038/srep31770} {\bibfield  {journal} {\bibinfo  {journal} {Scientific
  Reports}\ }\textbf {\bibinfo {volume} {6}},\ \bibinfo {pages} {srep31770}
  (\bibinfo {year} {2016})}\BibitemShut {NoStop}%
\bibitem [{\citenamefont {Ataie}\ \emph {et~al.}(2013)\citenamefont {Ataie},
  \citenamefont {Kuo}, \citenamefont {Wiberg}, \citenamefont {Tong},
  \citenamefont {Huynh}, \citenamefont {Alic},\ and\ \citenamefont
  {Radic}}]{Ataie2013}%
  \BibitemOpen
  \bibfield  {author} {\bibinfo {author} {\bibfnamefont {V.}~\bibnamefont
  {Ataie}}, \bibinfo {author} {\bibfnamefont {P.~P.}\ \bibnamefont {Kuo}},
  \bibinfo {author} {\bibfnamefont {A.}~\bibnamefont {Wiberg}}, \bibinfo
  {author} {\bibfnamefont {Z.}~\bibnamefont {Tong}}, \bibinfo {author}
  {\bibfnamefont {C.}~\bibnamefont {Huynh}}, \bibinfo {author} {\bibfnamefont
  {N.}~\bibnamefont {Alic}}, \ and\ \bibinfo {author} {\bibfnamefont
  {S.}~\bibnamefont {Radic}},\ }\href {\doibase 10.1364/OFC.2013.OTh3D.2}
  {\bibfield  {journal} {\bibinfo  {journal} {Optical Fiber Communication
  Conference/National Fiber Optic Engineers Conference 2013}\ ,\ \bibinfo
  {pages} {OTh3D.2}} (\bibinfo {year} {2013})}\BibitemShut {NoStop}%
\bibitem [{\citenamefont {Doylend}\ \emph {et~al.}(2011)\citenamefont
  {Doylend}, \citenamefont {Heck}, \citenamefont {Bovington}, \citenamefont
  {Peters}, \citenamefont {Coldren},\ and\ \citenamefont
  {Bowers}}]{doylend20112Dbeamsteering}%
  \BibitemOpen
  \bibfield  {author} {\bibinfo {author} {\bibfnamefont {J.~K.}\ \bibnamefont
  {Doylend}}, \bibinfo {author} {\bibfnamefont {M.}~\bibnamefont {Heck}},
  \bibinfo {author} {\bibfnamefont {J.~T.}\ \bibnamefont {Bovington}}, \bibinfo
  {author} {\bibfnamefont {J.~D.}\ \bibnamefont {Peters}}, \bibinfo {author}
  {\bibfnamefont {L.}~\bibnamefont {Coldren}}, \ and\ \bibinfo {author}
  {\bibfnamefont {J.}~\bibnamefont {Bowers}},\ }\href@noop {} {\bibfield
  {journal} {\bibinfo  {journal} {Optics Express}\ }\textbf {\bibinfo {volume}
  {19}},\ \bibinfo {pages} {21595} (\bibinfo {year} {2011})}\BibitemShut
  {NoStop}%
\bibitem [{\citenamefont {Sun}\ \emph {et~al.}(2013)\citenamefont {Sun},
  \citenamefont {Timurdogan}, \citenamefont {Yaacobi}, \citenamefont
  {Hosseini},\ and\ \citenamefont {Watts}}]{sun2013nanophotonicPhasedArrray}%
  \BibitemOpen
  \bibfield  {author} {\bibinfo {author} {\bibfnamefont {J.}~\bibnamefont
  {Sun}}, \bibinfo {author} {\bibfnamefont {E.}~\bibnamefont {Timurdogan}},
  \bibinfo {author} {\bibfnamefont {A.}~\bibnamefont {Yaacobi}}, \bibinfo
  {author} {\bibfnamefont {E.~S.}\ \bibnamefont {Hosseini}}, \ and\ \bibinfo
  {author} {\bibfnamefont {M.~R.}\ \bibnamefont {Watts}},\ }\href@noop {}
  {\bibfield  {journal} {\bibinfo  {journal} {Nature}\ }\textbf {\bibinfo
  {volume} {493}},\ \bibinfo {pages} {195} (\bibinfo {year}
  {2013})}\BibitemShut {NoStop}%
\bibitem [{\citenamefont {Hulme}\ \emph {et~al.}(2015)\citenamefont {Hulme},
  \citenamefont {Doylend}, \citenamefont {Heck}, \citenamefont {Peters},
  \citenamefont {Davenport}, \citenamefont {Bovington}, \citenamefont
  {Coldren},\ and\ \citenamefont {Bowers}}]{Hulme2015}%
  \BibitemOpen
  \bibfield  {author} {\bibinfo {author} {\bibfnamefont {J.~C.}\ \bibnamefont
  {Hulme}}, \bibinfo {author} {\bibfnamefont {J.~K.}\ \bibnamefont {Doylend}},
  \bibinfo {author} {\bibfnamefont {M.~J.~R.}\ \bibnamefont {Heck}}, \bibinfo
  {author} {\bibfnamefont {J.~D.}\ \bibnamefont {Peters}}, \bibinfo {author}
  {\bibfnamefont {M.~L.}\ \bibnamefont {Davenport}}, \bibinfo {author}
  {\bibfnamefont {J.~T.}\ \bibnamefont {Bovington}}, \bibinfo {author}
  {\bibfnamefont {L.~A.}\ \bibnamefont {Coldren}}, \ and\ \bibinfo {author}
  {\bibfnamefont {J.~E.}\ \bibnamefont {Bowers}},\ }\href {\doibase
  10.1364/OE.23.005861} {\bibfield  {journal} {\bibinfo  {journal} {Optics
  Express}\ }\textbf {\bibinfo {volume} {23}},\ \bibinfo {pages} {5861}
  (\bibinfo {year} {2015})}\BibitemShut {NoStop}%
\bibitem [{\citenamefont {Akhmediev}\ and\ \citenamefont
  {Ankiewicz}(2005)}]{akhmediev2005dissipative}%
  \BibitemOpen
  \bibfield  {author} {\bibinfo {author} {\bibfnamefont {N.}~\bibnamefont
  {Akhmediev}}\ and\ \bibinfo {author} {\bibfnamefont {A.}~\bibnamefont
  {Ankiewicz}},\ }\href {https://books.google.de/books?id=IiTTn0iZCOYC} {\emph
  {\bibinfo {title} {Dissipative Solitons}}},\ Lecture Notes in Physics\
  (\bibinfo  {publisher} {Springer Berlin Heidelberg},\ \bibinfo {year}
  {2005})\BibitemShut {NoStop}%
\bibitem [{\citenamefont {Herr}\ \emph {et~al.}(2013)\citenamefont {Herr},
  \citenamefont {Brasch}, \citenamefont {Jost}, \citenamefont {Wang},
  \citenamefont {Kondratiev}, \citenamefont {Gorodetsky},\ and\ \citenamefont
  {Kippenberg}}]{Herr2013}%
  \BibitemOpen
  \bibfield  {author} {\bibinfo {author} {\bibfnamefont {T.}~\bibnamefont
  {Herr}}, \bibinfo {author} {\bibfnamefont {V.}~\bibnamefont {Brasch}},
  \bibinfo {author} {\bibfnamefont {J.~D.}\ \bibnamefont {Jost}}, \bibinfo
  {author} {\bibfnamefont {C.~Y.}\ \bibnamefont {Wang}}, \bibinfo {author}
  {\bibfnamefont {N.~M.}\ \bibnamefont {Kondratiev}}, \bibinfo {author}
  {\bibfnamefont {M.~L.}\ \bibnamefont {Gorodetsky}}, \ and\ \bibinfo {author}
  {\bibfnamefont {T.~J.}\ \bibnamefont {Kippenberg}},\ }\href {\doibase
  10.1109/CLEOE-IQEC.2013.6801769} {\bibfield  {journal} {\bibinfo  {journal}
  {Nature Photonics}\ }\textbf {\bibinfo {volume} {8}},\ \bibinfo {pages} {145}
  (\bibinfo {year} {2013})},\ \Eprint {http://arxiv.org/abs/1211.0733}
  {1211.0733} \BibitemShut {NoStop}%
\bibitem [{\citenamefont {Suh}\ \emph {et~al.}(2016)\citenamefont {Suh},
  \citenamefont {Yang}, \citenamefont {Yang}, \citenamefont {Yi},\ and\
  \citenamefont {Vahala}}]{dualcomb2016vahala}%
  \BibitemOpen
  \bibfield  {author} {\bibinfo {author} {\bibfnamefont {M.-G.}\ \bibnamefont
  {Suh}}, \bibinfo {author} {\bibfnamefont {Q.-F.}\ \bibnamefont {Yang}},
  \bibinfo {author} {\bibfnamefont {K.~Y.}\ \bibnamefont {Yang}}, \bibinfo
  {author} {\bibfnamefont {X.}~\bibnamefont {Yi}}, \ and\ \bibinfo {author}
  {\bibfnamefont {K.~J.}\ \bibnamefont {Vahala}},\ }\href@noop {} {\bibfield
  {journal} {\bibinfo  {journal} {Science}\ }\textbf {\bibinfo {volume}
  {354}},\ \bibinfo {pages} {600} (\bibinfo {year} {2016})}\BibitemShut
  {NoStop}%
\bibitem [{\citenamefont {Yu}\ \emph {et~al.}(2016)\citenamefont {Yu},
  \citenamefont {Okawachi}, \citenamefont {Griffith}, \citenamefont
  {Picqu{\'e}}, \citenamefont {Lipson},\ and\ \citenamefont
  {Gaeta}}]{yu2016DKSdualcombMIR}%
  \BibitemOpen
  \bibfield  {author} {\bibinfo {author} {\bibfnamefont {M.}~\bibnamefont
  {Yu}}, \bibinfo {author} {\bibfnamefont {Y.}~\bibnamefont {Okawachi}},
  \bibinfo {author} {\bibfnamefont {A.~G.}\ \bibnamefont {Griffith}}, \bibinfo
  {author} {\bibfnamefont {N.}~\bibnamefont {Picqu{\'e}}}, \bibinfo {author}
  {\bibfnamefont {M.}~\bibnamefont {Lipson}}, \ and\ \bibinfo {author}
  {\bibfnamefont {A.~L.}\ \bibnamefont {Gaeta}},\ }\href@noop {} {\bibfield
  {journal} {\bibinfo  {journal} {arXiv:1610.01121}\ } (\bibinfo {year}
  {2016})}\BibitemShut {NoStop}%
\bibitem [{\citenamefont {Marin-Palomo}\ \emph {et~al.}(2016)\citenamefont
  {Marin-Palomo}, \citenamefont {Kemal}, \citenamefont {Karpov}, \citenamefont
  {Kordts}, \citenamefont {Pfeifle}, \citenamefont {Pfeiffer}, \citenamefont
  {Trocha}, \citenamefont {Wolf}, \citenamefont {Brasch}, \citenamefont
  {Anderson}, \citenamefont {Rosenberger}, \citenamefont {Vijayan},
  \citenamefont {Freude}, \citenamefont {Kippenberg},\ and\ \citenamefont
  {Koos}}]{Marin-Palomo2016}%
  \BibitemOpen
  \bibfield  {author} {\bibinfo {author} {\bibfnamefont {P.}~\bibnamefont
  {Marin-Palomo}}, \bibinfo {author} {\bibfnamefont {J.~N.}\ \bibnamefont
  {Kemal}}, \bibinfo {author} {\bibfnamefont {M.}~\bibnamefont {Karpov}},
  \bibinfo {author} {\bibfnamefont {A.}~\bibnamefont {Kordts}}, \bibinfo
  {author} {\bibfnamefont {J.}~\bibnamefont {Pfeifle}}, \bibinfo {author}
  {\bibfnamefont {M.~H.~P.}\ \bibnamefont {Pfeiffer}}, \bibinfo {author}
  {\bibfnamefont {P.}~\bibnamefont {Trocha}}, \bibinfo {author} {\bibfnamefont
  {S.}~\bibnamefont {Wolf}}, \bibinfo {author} {\bibfnamefont {V.}~\bibnamefont
  {Brasch}}, \bibinfo {author} {\bibfnamefont {M.~H.}\ \bibnamefont
  {Anderson}}, \bibinfo {author} {\bibfnamefont {R.}~\bibnamefont
  {Rosenberger}}, \bibinfo {author} {\bibfnamefont {K.}~\bibnamefont
  {Vijayan}}, \bibinfo {author} {\bibfnamefont {W.}~\bibnamefont {Freude}},
  \bibinfo {author} {\bibfnamefont {T.~J.}\ \bibnamefont {Kippenberg}}, \ and\
  \bibinfo {author} {\bibfnamefont {C.}~\bibnamefont {Koos}},\ }\href {\doibase
  10.1038/nature22387} {\bibfield  {journal} {\bibinfo  {journal} {Nature}\
  }\textbf {\bibinfo {volume} {546}},\ \bibinfo {pages} {274} (\bibinfo {year}
  {2016})}\BibitemShut {NoStop}%
\bibitem [{\citenamefont {Brasch}\ \emph {et~al.}(2017)\citenamefont {Brasch},
  \citenamefont {Lucas}, \citenamefont {Jost}, \citenamefont {Geiselmann},\
  and\ \citenamefont {Kippenberg}}]{brasch2017DKSselfreferencing}%
  \BibitemOpen
  \bibfield  {author} {\bibinfo {author} {\bibfnamefont {V.}~\bibnamefont
  {Brasch}}, \bibinfo {author} {\bibfnamefont {E.}~\bibnamefont {Lucas}},
  \bibinfo {author} {\bibfnamefont {J.~D.}\ \bibnamefont {Jost}}, \bibinfo
  {author} {\bibfnamefont {M.}~\bibnamefont {Geiselmann}}, \ and\ \bibinfo
  {author} {\bibfnamefont {T.~J.}\ \bibnamefont {Kippenberg}},\ }\href@noop {}
  {\bibfield  {journal} {\bibinfo  {journal} {Light: Science \& Applications}\
  }\textbf {\bibinfo {volume} {6}},\ \bibinfo {pages} {e16202} (\bibinfo {year}
  {2017})}\BibitemShut {NoStop}%
\bibitem [{\citenamefont {Jost}\ \emph {et~al.}(2015)\citenamefont {Jost},
  \citenamefont {Herr}, \citenamefont {Lecaplain}, \citenamefont {Brasch},
  \citenamefont {Pfeiffer},\ and\ \citenamefont {Kippenberg}}]{Jost2014link}%
  \BibitemOpen
  \bibfield  {author} {\bibinfo {author} {\bibfnamefont {J.~D.}\ \bibnamefont
  {Jost}}, \bibinfo {author} {\bibfnamefont {T.}~\bibnamefont {Herr}}, \bibinfo
  {author} {\bibfnamefont {C.}~\bibnamefont {Lecaplain}}, \bibinfo {author}
  {\bibfnamefont {V.}~\bibnamefont {Brasch}}, \bibinfo {author} {\bibfnamefont
  {M.~H.~P.}\ \bibnamefont {Pfeiffer}}, \ and\ \bibinfo {author} {\bibfnamefont
  {T.~J.}\ \bibnamefont {Kippenberg}},\ }\href {\doibase
  10.1364/OPTICA.2.000706} {\bibfield  {journal} {\bibinfo  {journal} {Optica}\
  }\textbf {\bibinfo {volume} {2}},\ \bibinfo {pages} {706} (\bibinfo {year}
  {2015})}\BibitemShut {NoStop}%
\bibitem [{\citenamefont {Pfeiffer}\ \emph {et~al.}(2016)\citenamefont
  {Pfeiffer}, \citenamefont {Kordts}, \citenamefont {Brasch}, \citenamefont
  {Zervas}, \citenamefont {Geiselmann}, \citenamefont {Jost},\ and\
  \citenamefont {Kippenberg}}]{Pfeiffer2015Damascene}%
  \BibitemOpen
  \bibfield  {author} {\bibinfo {author} {\bibfnamefont {M.~H.~P.}\
  \bibnamefont {Pfeiffer}}, \bibinfo {author} {\bibfnamefont {A.}~\bibnamefont
  {Kordts}}, \bibinfo {author} {\bibfnamefont {V.}~\bibnamefont {Brasch}},
  \bibinfo {author} {\bibfnamefont {M.}~\bibnamefont {Zervas}}, \bibinfo
  {author} {\bibfnamefont {M.}~\bibnamefont {Geiselmann}}, \bibinfo {author}
  {\bibfnamefont {J.~D.}\ \bibnamefont {Jost}}, \ and\ \bibinfo {author}
  {\bibfnamefont {T.~J.}\ \bibnamefont {Kippenberg}},\ }\href {\doibase
  10.1364/OPTICA.3.000020} {\bibfield  {journal} {\bibinfo  {journal} {Optica}\
  }\textbf {\bibinfo {volume} {3}},\ \bibinfo {pages} {20} (\bibinfo {year}
  {2016})}\BibitemShut {NoStop}%
\bibitem [{\citenamefont {Brasch}\ \emph {et~al.}(2016)\citenamefont {Brasch},
  \citenamefont {Geiselmann}, \citenamefont {Herr}, \citenamefont {Lihachev},
  \citenamefont {Pfeiffer}, \citenamefont {Gorodetsky},\ and\ \citenamefont
  {Kippenberg}}]{Brasch2014Cherenkov}%
  \BibitemOpen
  \bibfield  {author} {\bibinfo {author} {\bibfnamefont {V.}~\bibnamefont
  {Brasch}}, \bibinfo {author} {\bibfnamefont {M.}~\bibnamefont {Geiselmann}},
  \bibinfo {author} {\bibfnamefont {T.}~\bibnamefont {Herr}}, \bibinfo {author}
  {\bibfnamefont {G.}~\bibnamefont {Lihachev}}, \bibinfo {author}
  {\bibfnamefont {M.~H.~P.}\ \bibnamefont {Pfeiffer}}, \bibinfo {author}
  {\bibfnamefont {M.~L.}\ \bibnamefont {Gorodetsky}}, \ and\ \bibinfo {author}
  {\bibfnamefont {T.~J.}\ \bibnamefont {Kippenberg}},\ }\href
  {http://dx.doi.org/10.1126/science.aad4811} {\bibfield  {journal} {\bibinfo
  {journal} {Science}\ }\textbf {\bibinfo {volume} {351}},\ \bibinfo {pages}
  {357} (\bibinfo {year} {2016})}\BibitemShut {NoStop}%
\bibitem [{\citenamefont {Lindenmann}\ \emph {et~al.}(2012)\citenamefont
  {Lindenmann}, \citenamefont {Balthasar}, \citenamefont {Hillerkuss},
  \citenamefont {Schmogrow}, \citenamefont {Jordan}, \citenamefont {Leuthold},
  \citenamefont {Freude},\ and\ \citenamefont {Koos}}]{Lindenmann2012}%
  \BibitemOpen
  \bibfield  {author} {\bibinfo {author} {\bibfnamefont {N.}~\bibnamefont
  {Lindenmann}}, \bibinfo {author} {\bibfnamefont {G.}~\bibnamefont
  {Balthasar}}, \bibinfo {author} {\bibfnamefont {D.}~\bibnamefont
  {Hillerkuss}}, \bibinfo {author} {\bibfnamefont {R.}~\bibnamefont
  {Schmogrow}}, \bibinfo {author} {\bibfnamefont {M.}~\bibnamefont {Jordan}},
  \bibinfo {author} {\bibfnamefont {J.}~\bibnamefont {Leuthold}}, \bibinfo
  {author} {\bibfnamefont {W.}~\bibnamefont {Freude}}, \ and\ \bibinfo {author}
  {\bibfnamefont {C.}~\bibnamefont {Koos}},\ }\href {\doibase
  10.1364/OE.20.017667} {\bibfield  {journal} {\bibinfo  {journal} {Optics
  Express}\ }\textbf {\bibinfo {volume} {20}},\ \bibinfo {pages} {17667}
  (\bibinfo {year} {2012})}\BibitemShut {NoStop}%
\bibitem [{\citenamefont {Weimann}\ \emph {et~al.}(2014)\citenamefont
  {Weimann}, \citenamefont {Lauermann}, \citenamefont {Fehrenbach},
  \citenamefont {Palmer}, \citenamefont {Hoeller}, \citenamefont {Freude},\
  and\ \citenamefont {Koos}}]{Weimann2014}%
  \BibitemOpen
  \bibfield  {author} {\bibinfo {author} {\bibfnamefont {C.}~\bibnamefont
  {Weimann}}, \bibinfo {author} {\bibfnamefont {M.}~\bibnamefont {Lauermann}},
  \bibinfo {author} {\bibfnamefont {T.}~\bibnamefont {Fehrenbach}}, \bibinfo
  {author} {\bibfnamefont {R.}~\bibnamefont {Palmer}}, \bibinfo {author}
  {\bibfnamefont {F.}~\bibnamefont {Hoeller}}, \bibinfo {author} {\bibfnamefont
  {W.}~\bibnamefont {Freude}}, \ and\ \bibinfo {author} {\bibfnamefont {C.~G.}\
  \bibnamefont {Koos}},\ }in\ \href@noop {} {\emph {\bibinfo {booktitle}
  {Conference on Lasers and Electro-Optics}}}\ (\bibinfo {organization}
  {Optical Society of America},\ \bibinfo {year} {2014})\ pp.\ \bibinfo {pages}
  {STh4O--3}\BibitemShut {NoStop}%
\bibitem [{\citenamefont {Billah}\ \emph {et~al.}(2017)\citenamefont {Billah},
  \citenamefont {Blaicher}, \citenamefont {Kemal}, \citenamefont {Hoose},
  \citenamefont {Zwickel}, \citenamefont {Dietrich}, \citenamefont {Troppenz},
  \citenamefont {M{\"o}hrle}, \citenamefont {Merget}, \citenamefont {Hofmann}
  \emph {et~al.}}]{Billah2017}%
  \BibitemOpen
  \bibfield  {author} {\bibinfo {author} {\bibfnamefont {M.~R.}\ \bibnamefont
  {Billah}}, \bibinfo {author} {\bibfnamefont {M.}~\bibnamefont {Blaicher}},
  \bibinfo {author} {\bibfnamefont {J.~N.}\ \bibnamefont {Kemal}}, \bibinfo
  {author} {\bibfnamefont {T.}~\bibnamefont {Hoose}}, \bibinfo {author}
  {\bibfnamefont {H.}~\bibnamefont {Zwickel}}, \bibinfo {author} {\bibfnamefont
  {P.-I.}\ \bibnamefont {Dietrich}}, \bibinfo {author} {\bibfnamefont
  {U.}~\bibnamefont {Troppenz}}, \bibinfo {author} {\bibfnamefont
  {M.}~\bibnamefont {M{\"o}hrle}}, \bibinfo {author} {\bibfnamefont
  {F.}~\bibnamefont {Merget}}, \bibinfo {author} {\bibfnamefont
  {A.}~\bibnamefont {Hofmann}},  \emph {et~al.},\ }in\ \href@noop {} {\emph
  {\bibinfo {booktitle} {Optical Fiber Communication Conference}}}\ (\bibinfo
  {organization} {Optical Society of America},\ \bibinfo {year} {2017})\ pp.\
  \bibinfo {pages} {Th5D--6}\BibitemShut {NoStop}%
\bibitem [{\citenamefont {Dietrich}\ \emph {et~al.}(2016)\citenamefont
  {Dietrich}, \citenamefont {Reuter}, \citenamefont {Blaicher}, \citenamefont
  {Schneider}, \citenamefont {Billah}, \citenamefont {Hoose}, \citenamefont
  {Hofmann}, \citenamefont {Caer}, \citenamefont {Dangel}, \citenamefont
  {Offrein}, \citenamefont {M{\"{o}}hrle}, \citenamefont {Troppenz},
  \citenamefont {Zander}, \citenamefont {Freude},\ and\ \citenamefont
  {Koos}}]{Dietrich2016}%
  \BibitemOpen
  \bibfield  {author} {\bibinfo {author} {\bibfnamefont {P.-I.~C.}\
  \bibnamefont {Dietrich}}, \bibinfo {author} {\bibfnamefont {I.}~\bibnamefont
  {Reuter}}, \bibinfo {author} {\bibfnamefont {M.}~\bibnamefont {Blaicher}},
  \bibinfo {author} {\bibfnamefont {S.}~\bibnamefont {Schneider}}, \bibinfo
  {author} {\bibfnamefont {M.~R.}\ \bibnamefont {Billah}}, \bibinfo {author}
  {\bibfnamefont {T.}~\bibnamefont {Hoose}}, \bibinfo {author} {\bibfnamefont
  {A.}~\bibnamefont {Hofmann}}, \bibinfo {author} {\bibfnamefont
  {C.}~\bibnamefont {Caer}}, \bibinfo {author} {\bibfnamefont {R.}~\bibnamefont
  {Dangel}}, \bibinfo {author} {\bibfnamefont {B.}~\bibnamefont {Offrein}},
  \bibinfo {author} {\bibfnamefont {M.}~\bibnamefont {M{\"{o}}hrle}}, \bibinfo
  {author} {\bibfnamefont {U.}~\bibnamefont {Troppenz}}, \bibinfo {author}
  {\bibfnamefont {M.}~\bibnamefont {Zander}}, \bibinfo {author} {\bibfnamefont
  {W.}~\bibnamefont {Freude}}, \ and\ \bibinfo {author} {\bibfnamefont
  {C.}~\bibnamefont {Koos}},\ }\href {\doibase 10.1364/CLEO_SI.2016.SM1G.4}
  {\bibfield  {journal} {\bibinfo  {journal} {Conference on Lasers and
  Electro-Optics}\ ,\ \bibinfo {pages} {SM1G.4}} (\bibinfo {year}
  {2016})}\BibitemShut {NoStop}%
\bibitem [{\citenamefont {Roelkens}\ \emph {et~al.}(2010)\citenamefont
  {Roelkens}, \citenamefont {Liu}, \citenamefont {Liang}, \citenamefont
  {Jones}, \citenamefont {Fang}, \citenamefont {Koch},\ and\ \citenamefont
  {Bowers}}]{Roelkens2010}%
  \BibitemOpen
  \bibfield  {author} {\bibinfo {author} {\bibfnamefont {G.}~\bibnamefont
  {Roelkens}}, \bibinfo {author} {\bibfnamefont {L.}~\bibnamefont {Liu}},
  \bibinfo {author} {\bibfnamefont {D.}~\bibnamefont {Liang}}, \bibinfo
  {author} {\bibfnamefont {R.}~\bibnamefont {Jones}}, \bibinfo {author}
  {\bibfnamefont {A.}~\bibnamefont {Fang}}, \bibinfo {author} {\bibfnamefont
  {B.}~\bibnamefont {Koch}}, \ and\ \bibinfo {author} {\bibfnamefont
  {J.}~\bibnamefont {Bowers}},\ }\href {\doibase 10.1002/lpor.200900033}
  {\bibfield  {journal} {\bibinfo  {journal} {Laser \& Photonics Reviews}\
  }\textbf {\bibinfo {volume} {4}},\ \bibinfo {pages} {751} (\bibinfo {year}
  {2010})}\BibitemShut {NoStop}%
\bibitem [{\citenamefont {Joshi}\ \emph {et~al.}(2017)\citenamefont {Joshi},
  \citenamefont {Okawachi}, \citenamefont {Yu}, \citenamefont {Klenner},
  \citenamefont {Xingchen}, \citenamefont {Luke}, \citenamefont {Lipson},\ and\
  \citenamefont {Gaeta}}]{Joshi2017}%
  \BibitemOpen
  \bibfield  {author} {\bibinfo {author} {\bibfnamefont {C.~S.}\ \bibnamefont
  {Joshi}}, \bibinfo {author} {\bibfnamefont {Y.}~\bibnamefont {Okawachi}},
  \bibinfo {author} {\bibfnamefont {M.}~\bibnamefont {Yu}}, \bibinfo {author}
  {\bibfnamefont {A.}~\bibnamefont {Klenner}}, \bibinfo {author} {\bibfnamefont
  {J.}~\bibnamefont {Xingchen}}, \bibinfo {author} {\bibfnamefont
  {K.}~\bibnamefont {Luke}}, \bibinfo {author} {\bibfnamefont {M.}~\bibnamefont
  {Lipson}}, \ and\ \bibinfo {author} {\bibfnamefont {A.~L.}\ \bibnamefont
  {Gaeta}},\ }\href {\doibase 10.1038/nphys3875.37.} {\bibfield  {journal}
  {\bibinfo  {journal} {Conference on Lasers and Electro-Optics}\ ,\ \bibinfo
  {pages} {FTh4D.2}} (\bibinfo {year} {2017})},\ \Eprint
  {http://arxiv.org/abs/1610.01121} {arXiv:1610.01121} \BibitemShut {NoStop}%
\bibitem [{\citenamefont {Guo}\ \emph {et~al.}(2017)\citenamefont {Guo},
  \citenamefont {Karpov}, \citenamefont {Lucas}, \citenamefont {Kordts},
  \citenamefont {Pfeiffer}, \citenamefont {Lichachev}, \citenamefont {Lobanov},
  \citenamefont {Gorodetsky},\ and\ \citenamefont {Kippenberg}}]{Karpov2016}%
  \BibitemOpen
  \bibfield  {author} {\bibinfo {author} {\bibfnamefont {H.}~\bibnamefont
  {Guo}}, \bibinfo {author} {\bibfnamefont {M.}~\bibnamefont {Karpov}},
  \bibinfo {author} {\bibfnamefont {E.}~\bibnamefont {Lucas}}, \bibinfo
  {author} {\bibfnamefont {A.}~\bibnamefont {Kordts}}, \bibinfo {author}
  {\bibfnamefont {M.~H.~P.}\ \bibnamefont {Pfeiffer}}, \bibinfo {author}
  {\bibfnamefont {G.}~\bibnamefont {Lichachev}}, \bibinfo {author}
  {\bibfnamefont {V.~E.}\ \bibnamefont {Lobanov}}, \bibinfo {author}
  {\bibfnamefont {M.~L.}\ \bibnamefont {Gorodetsky}}, \ and\ \bibinfo {author}
  {\bibfnamefont {T.~J.}\ \bibnamefont {Kippenberg}},\ }\href
  {https://www.nature.com/nphys/journal/v13/n1/full/nphys3893.html} {\bibfield
  {journal} {\bibinfo  {journal} {Nature Physics}\ }\textbf {\bibinfo {volume}
  {13}},\ \bibinfo {pages} {94} (\bibinfo {year} {2017})}\BibitemShut {NoStop}%
\bibitem [{\citenamefont {Allan}(1966)}]{Allan.1966}%
  \BibitemOpen
  \bibfield  {author} {\bibinfo {author} {\bibfnamefont {D.~W.}\ \bibnamefont
  {Allan}},\ }\href {\doibase 10.1109/PROC.1966.4634} {\bibfield  {journal}
  {\bibinfo  {journal} {Proceedings of the IEEE}\ }\textbf {\bibinfo {volume}
  {54}},\ \bibinfo {pages} {221} (\bibinfo {year} {1966})}\BibitemShut
  {NoStop}%
\bibitem [{pip()}]{pipositioner}%
  \BibitemOpen
  \href@noop {} {{\selectlanguage {English}\emph {\bibinfo {title} {MP84E User
  Manual M-511.HD Ulta-High-Resolution Positioner}}}},\ \bibinfo {organization}
  {Physics Instruments}\BibitemShut {NoStop}%
\bibitem [{\citenamefont {Ciddor}(1996)}]{Ciddor.1996}%
  \BibitemOpen
  \bibfield  {author} {\bibinfo {author} {\bibfnamefont {P.~E.}\ \bibnamefont
  {Ciddor}},\ }\href {\doibase 10.1364/AO.35.001566} {\bibfield  {journal}
  {\bibinfo  {journal} {Applied Optics}\ }\textbf {\bibinfo {volume} {35}},\
  \bibinfo {pages} {1566} (\bibinfo {year} {1996})}\BibitemShut {NoStop}%
\bibitem [{\citenamefont {Suh}\ and\ \citenamefont
  {Vahala}(2017)}]{suh2017VahalaLIDAR}%
  \BibitemOpen
  \bibfield  {author} {\bibinfo {author} {\bibfnamefont {M.-G.}\ \bibnamefont
  {Suh}}\ and\ \bibinfo {author} {\bibfnamefont {K.}~\bibnamefont {Vahala}},\
  }\href@noop {} {\bibfield  {journal} {\bibinfo  {journal} {arXiv:1705.06697}\
  } (\bibinfo {year} {2017})}\BibitemShut {NoStop}%
\end{thebibliography}

------------------

%\includepdf[pages={1}]{Figure2_PT_v11mk.pdf}

\end{document}